\documentclass[showpacs,amsmath,amssymb,amsthm,aps,jmp,thmsb]{revtex4}

\usepackage{color}

\usepackage{amsmath}
\usepackage{latexsym}
\usepackage{amsfonts}
\usepackage{amssymb}

\newtheorem{proposition}{Proposition}
\newtheorem{theorem}{Theorem}
\newtheorem{lemma}{Lemma}
\newtheorem{corollary}{Corollary}

\newenvironment{proof}[1][Proof]{\textbf{#1.} }{\ \rule{0.5em}{0.5em}}

\newcommand{\R}{\mathbb R} 
\newcommand{\N}{\mathbb N} 

\newcommand{\hi}{\mathcal H} 
\newcommand{\ip}[2]{\langle {#1}|{#2}\rangle} 
\newcommand{\tr}[1]{\mathrm{tr}\left[ {#1} \right]} 
\newcommand{\fii}{\varphi}
\newcommand{\veps}{\varepsilon}

\newcommand{\sfq}{\mathsf{Q}}
\newcommand{\sfp}{\mathsf{P}}
\newcommand{\qhat}{Q}
\newcommand{\phat}{P}

\newcommand{\br}{\mathcal{B}(\R)} 
\newcommand{\brr}{\mathcal{B}(\R^2)} 
\newcommand{\prob}{\mathsf{p}} 

\newcommand{\W}{{\mathcal W}}

\newcommand{\delq}[1]{\W_{\varepsilon_1,\delta}({#1},\sfq)}
\newcommand{\delp}[1]{\W_{\varepsilon_2,\delta}({#1},\sfp)}
\newcommand{\deq}[1]{\W_{\varepsilon_1}({#1},\sfq)}
\newcommand{\dep}[1]{\W_{\varepsilon_2}({#1},\sfp)}

\newcommand{\dele}[1]{\mathcal{W}_{\varepsilon,\delta}({#1},E)}
\newcommand{\dee}[1]{\mathcal{W}_{\varepsilon}({#1},E)}

\newcommand{\mav}{M^{\mathrm av}}
\newcommand{\bom}{{\mathbf m}}
\newcommand{\iqd}{J_{q;\delta}}
\newcommand{\ixd}{J_{x;\delta}}
\newcommand{\ipd}{J_{p;\delta}}
\newcommand{\iqw}{J_{q;w}}

\newcommand{\ipw}{J_{p;w'}}
\newcommand{\intv}[2]{J_{{#1};{#2}}}

\begin{document}

\title
{Universal joint-measurement uncertainty relation for error bars}

\author{P. Busch}
\email{pb516@york.ac.uk} \affiliation{\noindent\hbox{Perimeter Institute for Theoretical
Physics, Waterloo, Canada,}\\ and Department of Mathematics, University of York,
York, UK}

\author{D.B. Pearson}
\email{d.b.pearson@hull.ac.uk} \affiliation{\noindent\hbox{Department of Mathematics, University of Hull,
Hull, UK}}

\date{20 April 2007 (small corrections: 22 June 2007)}

\begin{abstract}
\noindent 
We formulate and prove a new, universally valid uncertainty relation for the necessary errors bar widths in any approximate joint measurement of position and momentum.


\end{abstract}
\maketitle



\section{Introduction}\label{intro}

In his seminal paper of 1927 \cite{Heisenberg27}, Heisenberg
envisaged not one but in fact three conceptually distinct variants
of uncertainty relations for position and momentum of the general form
\begin{equation}\label{gen-ur}
\delta q\cdot\delta p\gtrsim h
\end{equation}
which together comprise the full content of the uncertainty: this relation
can be read as describing a trade-off  (a) between the widths of the probability distributions of
position and momentum in a quantum state; (b)  between the inaccuracies
of an approximate joint measurement; and (c) between the accuracy
of a measurement of (say) position and the ensuing unavoidable disturbance of the
momentum (distribution).

The latter two versions have until recently lacked a rigorous formal basis and their universal
validity has accordingly been questioned. Here we formulate and prove a form of the
joint-measurement uncertainty relation
(b) in terms of a new concept of \emph{error bar width}. In \cite{BuHeLa06} it is shown how
the inaccuracy-disturbance relation (c) arises as a consequence. Our proof is an
adaptation of a strategy recently developed by R.~Werner \cite{Werner04b} who proved
``uncertainty" relations in the spirit of (b) and (c) for a distance measure between observables.
In contrast to Werner's geometric measure of distance, our measure of error bar width is modeled
in close analogy to the experimental physicists' way of estimating errors. We will also show that
the notion of approximation in the sense of finite error bars is more general than that in terms of
finite distance.

\section{Approximate measurements and error bar width}\label{sec:errorbar}

 \subsection{Preliminaries}

Throughout the paper we consider a quantum particle in one
spatial dimension, with Hilbert space $\hi=L^2(\R)$ and canonical
position and momentum operators $\qhat,\phat$, defined
in the usual way via $(\qhat\psi)(x)=x\psi(x)$,
$(\phat\psi)(x)=-i\hbar (d\psi/dx)(x)$.
Generalizations to more degrees of freedom are straightforward.
By $\sfq$ and $\sfp$ we denote the
spectral measures of $\qhat$ and $\phat$, respectively, and
$W(q,p)=e^{\frac i{2\hbar} qp}\,e^{-\frac i\hbar q\phat}\,
e^{\frac i\hbar p\qhat}$ are the Weyl operators
which comprise an irreducible unitary projective representation of
the translations on phase space $\R^2$. States are represented as
positive operators $\rho$ of trace 1, the convex set of all states
being denoted  $S$. 

Observables are represented as normalized
($E(\Omega)=I$) positive operator measures (POMs) on a measurable
space $(\Omega,\Sigma)$, which in the present context will be one of
the Borel spaces $(\R,\br)$ or $(\R^2,\brr)$. An observable $E$ is called
\emph{sharp} if it is projection valued; otherwise $E$ is an \emph{unsharp}
observable. We write $\rho^E$ for
the probability measure induced by a state $\rho$ and an observable
$E$ via the formula $\rho^E(X)=\tr{\rho E(X)}$, $X\in \Sigma$.

The \emph{overall width} (at confidence level $1-\veps$) of a probability measure
$\prob$ on $\R$ is defined for $\veps\in[0,1)$ as
\begin{equation}
W_\veps(\prob):=\inf\{w>0\,|\,\exists x\in\R: \,\prob([x-\tfrac w2,x+\tfrac w2])\ge 1-\veps\}.
\end{equation}
Note that the overall width is finite for any $\veps>0$.

In analogy to the uncertainty relation for standard deviations,
the overall widths of the position
and momentum distributions in a state $\rho$ also satisfy a trade-off relation: for positive
$\veps_1,\veps_2>0$, the inequality
\begin{equation}\label{state-ow-ur}
W_{\veps_1}(\rho^\sfq)\cdot W_{\veps_2}(\rho^\sfp)\ge 2\pi\hbar\cdot(1-\veps_1-\veps_2)^2
\end{equation}
holds for all $\rho\in S$ if  $\veps_1+\veps_2<1$. (For $\veps_1+\veps_2\ge 1$ there is no
positive lower bound for the product on the left hand side.)
Uncertainty relations of this form have been obtained by various authors,
based on results of \cite{LaPo61}. The lower
bound given here was obtained in \cite{BuHeLa06} using a simple argument.
To our knowledge, the sharpest lower bound known so far is given by 
 Uffink in 1990 \cite{Uffink90}:
\begin{equation}
2\pi\hbar\cdot\left(\sqrt{(1-\veps_1)(1-\veps_2)}-\sqrt{\veps_1\veps_2}\right)^2.
\end{equation}
This term can be substituted for
$2\pi\hbar(1-\veps_1-\veps_2)^2$ here and in
all subsequent applications of (\ref{state-ow-ur}).

\subsection{Approximate joint  measurements}

A pair of observables $M_1,M_2$  on $\R$ is said to be \emph{jointly measurable} if
there is an observable $M$ on $\R^2$ of which
$M_1$, $M_2$ are the marginals ($M_1(X)=M(X\times\R), M_2(Y)=M(\R\times Y)$).
Observable $M$ is called a \emph{joint observable} for $M_1,M_2$.

It is a fundamental fact that pairs of sharp quantum observables are jointly
measurable exactly when they commute. However, there are pairs $M_1,M_2$ of
unsharp observables that are mutually noncommuting but do have a joint
observable. This opens up the general
possibility of defining an \emph{approximate joint measurement} of
two noncommuting observables $E_1,E_2$ as a joint measurement of
two observables $M_1,M_2$ which are approximations of $E_1,E_2$ in an
appropriate sense.
The deviation of $M_i$ from $E_i$
will be referred to as \emph{error} or  \emph{inaccuracy}.

The notion of an approximate joint measurements of two noncommuting observables
$E_1$ and $E_2$ draws thus on the idea of deliberately allowing inaccuracy and
intrinsic unsharpness, in the
hope that one can find approximations $M_1$ and $M_2$ to $E_1$ and $E_2$ which
arise as marginals of some observable $M$.
We will show that for any observable $M$ on phase space the marginals $M_1,M_2$
cannot be both arbitrarily good approximations  to $\sfq,\sfp$, respectively. If they
are to be approximations, they will also have to be sufficiently unsharp.


\subsection{Error bar width}

The following definition of an error measure is guided by the notion of calibrating a measuring
instrument by testing it with input states that represent sharp values of the quantity to be measured.
This procedure serves to estimate likely error bars.

For simplicity, we give our definitions  of  approximations only for sharp observables $E$ on
$\br$ which are supported on $\R$ (meaning here that $E(J)$ differs from the null operator $O$ for
any open interval $J$), so that the assumption of localized input states can be described as
$\rho^E(J_{x;\delta})=1$, for 
 any interval
 $J_{x;\delta}:=[x-\delta/2,x+\delta/2]$, $x\in\R,\delta>0$.

Let $E_1$ be an observable on $\R$.
For each $\varepsilon\in(0,1)$, $\delta>0$, we define the \emph{error} of $E_1$ relative to $E$
\begin{equation}
\dele{E_1}:=\inf\{w>0\,|\ \forall\ x\in\R\ \forall\rho\in S:\,\rho^E(J_{x;\delta})=1\Rightarrow
 \rho^{E_1}(J_{x,w})\ge 1-\veps\}.
\end{equation}
The error describes the range within which the input values can be inferred
from the output distributions, with confidence level $1-\varepsilon$, given initial
localizations within $\delta$.

We say that $E_1$ is an $\veps$-\emph{approximation} to $E$ if
$\dele{E_1}<\infty$ for all $\delta>0$.\footnote{The fact that this condition is required
for \emph{all} $\delta$ reflects the idea that calibrations at confidence level $1-\veps$
should be valid on all scales.}
We note that the error is an increasing function of $\delta$, so that we can
define the \emph{error bar width} of $E_1$ relative to $E$:
\begin{equation}
\dee{E_1}:=\inf_\delta\dele{E_1}=\lim_{\delta\to 0}\dele{E_1}.
\end{equation}
In case $\dele{E_1}=\infty$ for all $\delta>0$, we write $\dee{E_1}=\infty$. If $E_1=E$, then
$\dee{E_1}=0$ for all $\veps\in(0,1)$.

$E_1$ will be called an \emph{approximation} to $E$ if
$\dee{E_1}<\infty$ for all $\veps\in(0,1)$.\footnote{One will only
consider $E_1$ to be an approximation of $E$ if for a given input
distribution $\rho^E$ supported within $\ixd$ the output
distribution $\rho^{E_1}$ is \emph{concentrated} around the set
$\ixd$, that is, assigns probability greater than $1/2$ to some
interval $\intv{x}{w}$. This means that the definition of an
$\veps$-approximation is only of interest for $\veps\in(0,1/2)$.
However, since the quantity $\dee{E_1}$ is a decreasing function of
$\veps$, the defining condition would still be satisfied for all
$\veps\in(0,1)$ if it holds for $\veps\in(0,1/2)$.}

We say that an observable $M$ on $\R^2$ is an \emph{approximate joint observable} of $E_1,E_2$
if the marginals $M_1,M_2$ are approximations to $E_1,E_2$, respectively.

A detailed analysis of these definitions will be given elsewhere \cite{BuPe07}.

\subsection{Resolution width}

As an indicator of the intrinsic unsharpness of an observable $E_1$ on $\br$, we use the
\emph{resolution width} (at confidence level $1-\veps$),  defined as follows \cite{CaHeTo06}:
\begin{equation}
\gamma_\veps(E_1):=\inf\{ w>0\,|\, \forall x\in\R\,\exists\rho\in S: \ \rho^{E_1}(\intv{x}{w})\ge 1-\veps  \}.
\end{equation}
For a sharp observable $E$ on $\br$ with support $\R$ the resolution width is $\gamma_\veps(E)=0$
for all $\veps\in(0,1)$.
\begin{proposition}\label{err-res}
Let $E_1,E$ be observables on $\R$, and $E$ be sharp with support $\R$.
The error bar width of $E_1$ relative
to $E$ is never smaller than the resolution width of $E_1$:
\begin{equation}\label{w-gamma}
\dee{E_1}\ge \gamma_\veps(E_1).
\end{equation}
\end{proposition}

\begin{proof}
If $\dee{E_1}=\infty$, the inequality is trivially satisfied.
Assume that $\dee{E_1}$ is finite. There is a $\delta_0>0$ such that
$\W_{\veps,\delta_0}(E_1,E)<\infty$. Since $\dele{E_1}$ is an increasing function of $\delta$,
we also have $\dele{E_1}<\infty$ for $\delta\le\delta_0$. Let
$w\ge\dele{E_1}$ for some $\delta$, $0<\delta\le\delta_0$. Thus
for all $x\in\R$ and all $\rho$ with $\rho^E(\ixd)=1$ we have $\rho^{E_1}(\intv{x}{w})\ge 1-\veps$.
This entails (given that the support of $E$ is $\R$) that for all $x\in\R$ there is some $\rho$ such that
$\rho^{E_1}(\intv{x}{w})\ge 1-\veps$. Hence $w\ge\gamma_\veps(E_1)$, and therefore
$\dele{E_1}\ge\gamma_\veps(E_1)$ for all $\delta>0$, from which (\ref{w-gamma}) follows.
\end{proof}

\begin{corollary}
Let $E$ be an observable on $\br$ with support $\R$. Any $\veps$-approxima\-tion $E_1$ of $E$
has finite resolution width, $\gamma_{\veps}(E_1)<\infty$.
\end{corollary}

\section{Uncertainty relations for phase space observables}

\subsection{Approximate position and momentum}

An important class of candidates of approximate observables for
position and momentum  are obtained as smearings of $\sfq$ and $\sfp$, for example,
by means of convolutions with probability measures
$\mu,\nu$. Thus, observables
$\sfq_\mu,\sfp_\nu$  are defined via the weak integrals
\begin{equation}\label{approx-pos-mom}\begin{split}
\sfq_\mu(X)&=\sfq*\mu(X)=\int_\R\mu(X+q)\,\sfq(dq),\\
\sfp_\nu(Y)&=\sfp*\nu(Y)=\int_\R\nu(Y+p)\,\sfp(dp).
\end{split}
\end{equation}
These shift-covariant observables will be  called \emph{approximate position and momentum.}

\begin{proposition}\label{prop:qp-approx}
$\sfq_\mu$ and $\sfp_\nu$ are  approximations to $\sfq$ and $\sfp$ for
any probability measures $\mu$ and $\nu$, respectively.
\end{proposition}

\begin{proof}
It suffices to consider the case of $\sfq_\mu$.

Let $\veps\in(0,1), \delta>0$ be given.
Let $q_0,w_0$ be such that $\mu(J_{q_0;w_0})\ge 1-\veps$. Then, for
$w\ge 2|q_0|+w_0+\delta$, it follows that $\iqd\subseteq x+\iqw$ for all $x\in J_{q_0;w_0}$.

Now let $q\in\R$, and let $\rho\in S$ be such that $\rho^\sfq(\iqd)=1$. Then
$\rho^\sfq(x+\iqw)=1$ for all  $x\in J_{q_0;w_0}$, and therefore:
\begin{equation}\label{rho-ineq}\begin{split}
\rho^{\sfq_\mu}(\iqw)&=\int\mu(dx)\rho^\sfq(x+\iqw)\\
&\ge
\int_{J_{q_0;w_0}}\mu(dx)\rho^\sfq(x+\iqw)=\mu(J_{q_0;w_0})
\ge1-\veps.
\end{split}
\end{equation}
\end{proof}

\begin{proposition}
Observables $\sfq_\mu$ and $\sfp_\nu$ satisfy the following relations:
\begin{equation}\label{qp-err-res}
\deq{\sfq_\mu}\ge\gamma_{\veps_1}(\sfq_\mu)=W_{\veps_1}(\mu),\quad
\dep{\sfp_\nu}\ge\gamma_{\veps_2}(\sfp_\nu)=W_{\veps_2}(\nu).
\end{equation}
\end{proposition}

\begin{proof} The inequalities are a consequence of Proposition \ref{err-res}.
It remains to prove the equalities, which we will do for the case
$\gamma_{\veps_1}(\sfq_\mu)=W_{\veps_1}(\mu)$.

Assume a positive number $w$ is given such that $w\ge \gamma_{\veps_1}(\sfq_\mu)$. Thus,
 for any $q\in\R$ there is a state $\rho$ with
 \[
 \rho^{\sfq_\mu}(\iqw)=\int_\R \rho^\sfq(dq')\mu(\iqw+q')\ge1-\veps_1.
 \]
This shows that it is impossible to have $\mu(\iqw+q')<1-\veps_1$ for all $q'$, so that there exists
 a $q'$ with $\mu(\iqw+q')\ge1-\veps_1$. This means that $w\ge W_{\veps_1}(\mu)$.
Hence $\gamma_{\veps_1}(\sfq_\mu)\ge W_{\veps_1}(\mu)$.

To show the converse inequality, $W_{\veps_1}(\mu)\ge \gamma_{\veps_1}(\sfq_\mu)$,
let $w>W_{\veps_1}(\mu)$. Then there exists an interval $K$ of length $w$ such that 
$\mu(K)\ge 1-\veps_1$. Now let $J_q$ be any interval of length greater than $w$.  Since the length
of $J_q$ is greater than the length of $K$, it follows that the intersection of all intervals $J_q+x$,
as $x$ runs over $K$, is an interval of positive length. 
This interval, which is contained in $J_q+x$ for all $x\in K$, we denote by $J_0$. 

Let $\rho$ be any state concentrated in $J_0$, so that $\rho^\sfq(J_q+x)=1$ for  $x\in K$. From formula
(\ref{rho-ineq}), this gives $\rho^{\sfq_\mu}(J_q)\ge\mu(K)\ge 1-\veps_1$. Hence 
$w\ge\gamma_{\veps_1}(\sfq_\mu)$.
Since $w>W_{\veps_1}(\mu)$ was arbitrary, the required result follows.
\end{proof}

The question which pairs $\sfq_\mu,\sfp_\nu$ are jointly measurable has a complete answer,
proven in \cite{CaHeTo05}: they have to be marginals of a covariant phase space observable.

\subsection{Covariant phase space observables}

An observable $G$ on phase space $\R^2$ will be called a \emph{phase space observable}
if it satisfies the covariance condition
\begin{equation}\label{gen-G-covar}
W(q,p)G(Z)W(q,p)^*=G(Z+(q,p)).
\end{equation}
for all $Z\in\brr$.

It is known that all covariant phase space observables are of the form $G=G^\bom$,
\begin{equation}\label{gen-cov-obs}
\mathcal{B}(\R^2)\ni Z\mapsto G^\bom(Z)=   \frac 1{2\pi\hbar}\int_Z W(q,p)\bom W(q,p)^*dqdp,
\end{equation}
where  the integral is defined weakly and the operator density is
generated  by an arbitrary fixed positive operator $\bom$ of trace 1.
This fundamental fact has been proven and extensively studied by
several authors using different techniques \cite{PSAQT,Werner84,CaDeTo03,KiLaYl06a}.

The marginal observables of $G^\bom$ are of the form (\ref{approx-pos-mom}),
with the probability measures
$\mu_\bom:=\bom_\varPi^\sfq,\nu_\bom:=\bom_\varPi^\sfp$, that is,
$G^\bom_1=\sfq*\mu_\bom$, $G^\bom_2=\sfp*\nu_\bom$.
Here $\bom_\varPi=\varPi\bom\varPi^*$ is the operator obtained from $\bom$
under the action of the parity transformation $\varPi$
($\varPi\fii(x)=\fii(-x)$).

As shown in \cite{CaHeTo05}, observables $\sfq_\mu,\sfp_\nu$ are jointly measurable exactly when there is a {\em covariant} phase space observable $G^\bom$ of
they are the marginals.
 In that case the resolution widths are given by the widths of
the probability measures $\mu_\bom,\nu_\bom$ (via Eq.~(\ref{qp-err-res}))
which obey the uncertainty relation (\ref{state-ow-ur}); hence,
\begin{equation}\label{cov-res-ur}
\gamma_{\veps_1}(\sfq_{\mu_\bom})\cdot\gamma_{\veps_2}(\sfp_{\nu_\bom})=
W_{\veps_1}(\mu_\bom)\cdot W_{\veps_2}(\nu_\bom)\ge 2\pi\hbar\cdot
(1-\veps_1-\veps_2)^2
\end{equation}
for any $\veps_1,\veps_2>0$ with $\veps_1+\veps_2<1$.

\begin{proposition}\label{cov-obs-ur}
Any covariant phase space observable $G^\bom$ with generating density operator $\bom$
is an approximate
joint observable for $\sfq,\sfp$, with
the error bar widths satisfying the joint measurement
uncertainty relation
\begin{equation}\label{cov-error-bar-ur}
\deq{\sfq_{\mu_\bom}}\cdot\dep{\sfp_{\nu_\bom}}
\ge {2\pi\hbar}\cdot (1-\veps_1-\veps_2)^2
\end{equation}
for any $\veps_1,\veps_2>0$ with $\veps_1+\veps_2<1$.
\end{proposition}

\begin{proof} The first statement is a direct consequence of Proposition \ref{prop:qp-approx}. The inequality follows from Eqs.~(\ref{qp-err-res}) and (\ref{cov-res-ur}).
\end{proof}

\section{Uncertainty relations for general observables on phase space}

An observable $M$ on phase space $\R^2$ is an \emph{$(\veps_1,\veps_2)$-approximate}
joint observable of position and momentum if the marginal $M_1$ is
an $\veps_1$-approximation to $\sfq$ and the marginal 
 $M_2$
 is an $\veps_2$-approx\-imation to $\sfp$.
For later use we state this condition explicitly:
\\
\emph{For any $\delta>0$,
there are positive
numbers $w,w'<\infty$ such that the following conditions hold:
\begin{itemize}
\item[\rm{($\alpha$)}] for all $q\in\R$ and all $\rho\in\mathcal{S}$, if $\rho^\sfq(\iqd)=1$, then
$\rho^{M_1}(\iqw)\ge 1-\veps_1$;
\item[\rm{($\beta$)}] for all $p\in\R$ and all $\rho\in\mathcal{S}$, if $\rho^\sfp(\iqd)=1$, then
$\rho^{M_2}(\ipw)\ge 1- \veps_2$.
\end{itemize}
}

Our main result is the following.

\begin{theorem}\label{thm}
Let $M$ be an approximate joint observable for $\sfq,\sfp$. Then,
for $\veps_1,\veps_2\in(0,1)$ with $\veps_1+\veps_2<1$,
the error bar widths and resolutions widths of $M_1$ and $M_2$ satisfy the
uncertainty relations
\begin{eqnarray}
\deq{M_1}\cdot\dep{M_2}&\ge& 2\pi\hbar\cdot (1-\veps_1-\veps_2)^2,\label{error-bar-ur}
\nonumber\\
\gamma_{\veps_1}(M_1)\cdot\gamma_{\veps_2}(M_2)
&\ge&2\pi\hbar\cdot (1-\veps_1-\veps_2)^2.\label{res-ur}
\end{eqnarray}
\end{theorem}

The remainder of this section develops the proof of Theorem
\ref{thm}. The proof strategy is adapted from recent work of R.~Werner \cite{Werner04b}
who derived a Heisenberg uncertainty relation for approximate joint measurements
of position and momentum in terms of a distance measure between two
observables.

We set out to show that if  $M$  is an approximate
joint observable of $\sfq,\sfp$, there is a covariant phase space
observable $G^\bom$ whose resolutions are not worse than those of $M$,
that is,  $\delq{G^\bom_i}\le\delq{M_i}$, $i=1,2$.  The
uncertainty relation (\ref{error-bar-ur}) was already proven for $G^\bom$
in Proposition \ref{cov-obs-ur}.

Following \cite{Werner04b}, we make use of the concept of the invariant mean
on the  group of phase space translations to introduce a covariant
phase space observable $\mav$ associated with any observable $M$ on
phase space. The invariant mean is a positive linear functional $\eta$ on
$C(\R^2)$ with the invariance property
$\eta(\tau_xf)=\eta(f)$.
(Here $\tau_x$,   $x=(q,p)\in\R^2$,  is the shift map on the space
of bounded Borel functions $f$, so that
$\tau_xf(y)=f(y-x)$.)  This extends the operation of integrating
$f$ over an interval, dividing by the interval length, and letting
that length go to infinity. While this operation only works for a
very limited class of functions, the existence of $\eta$ is
guaranteed by the axiom of choice.

Any observable $M$ on phase space can be viewed as a linear map from
the space $C_{uc}(\R^2)$ of bounded uniformly continuous functions to
the bounded operators on $\hi$
via $M(f)=\int f(q,p)dM(q,p)$ 
\cite[Lemma 2]{Werner04b}.
The marginals $M_1,M_2$ can then equally
be defined with respect to functions $f,g\in C_{uc}(\R)$ since such function
can be extended to the functions $F,G\in C_{uc}(\R^2)$, where $F(q,p):=f(q)$,
$G(q,p):=g(p)$; then $M_1(f):=M(F)$ and $M_2(g):=M(G)$.

For a POM  $M$ on $\brr$, an associated linear map $\mav$ is
defined via the following equations, required
to hold for any $f\in C_{uc}(\R^2)$ and all $\rho\in S$:
\begin{equation}\begin{split}\label{inv-mean}
\tr{\rho\mav(f)}&=\eta(u(\rho,f)),\\
u(\rho,f)(q,p)&=\tr{W(q,p)\rho W(q,p)^*M(\tau_{(q,p)}f)}=:\tr{\rho M^{(q,p)}(f)}.
\end{split}
\end{equation}
The covariance of
$\mav$,
\begin{equation}
W(q,p)\mav(f)W(q,p)^*=\mav(\tau_{(q,p)}f),
\end{equation}
is an immediate consequence of the invariance of $\eta$.
The marginals $\mav_1,\mav_2$ are defined according to the
prescription given in the preceding paragraph.

In order to apply and check the conditions of an approximate joint
measurement  to $\mav$, we need to restate the definition in terms
of  $M(f)$, $f\in C_{uc}(\R^2)$. In fact, we only need to refer
to $M_1(f),M_2(g)$ with $f,g\in C_{uc}(\R)$. Let $\chi_J$ denote the
characteristic function of the set $J$.

\begin{lemma}
Let $\veps_1,\veps_2\in(0,\tfrac 12)$ be given. An observable $M$ on phase
space $\R^2$ is an $(\veps_1,\veps_2)$-approximate joint observable
for $\sfq,\sfp$ if and only if the following conditions hold: for
any $\delta>0$, there are positive finite numbers $w,w'$ such that:
\begin{quote}
\item[\rm{($\alpha^\prime$)}]  for all $q\in\R$, all  $f\in C_{uc}(\R)$ with $\chi_{\iqw}\le f\le 1$
and all $\rho$ with $\rho^\sfq(\iqd)=1$, one has $\rho^{ M_1}(f)\ge 1-\veps_1$;
\item[\rm{($\beta^\prime$)}] for all $p\in\R$, all  $g\in C_{uc}(\R)$ with $\chi_{\ipw}\le g\le 1$
and all $\rho$ with $\rho^\sfp(\ipd)=1$, one has $\rho^{ M_2}(g)\ge 1-\veps_2$.
\end{quote}
\end{lemma}

\begin{proof}
Assume that $M$ is an $(\veps_1,\veps_2)$-approximate
joint observable for $\sfq,\sfp$. For given $\delta$, there exist
$w,w'<\infty$ such that the conditions ($\alpha$), ($\beta$) (formulated
just before Theorem 1) hold. Then ($\alpha'$), ($\beta'$) follow
immediately since due to the monotonicity of $M_1$ we have
$\rho^{ M_1}(\iqw)\le\rho^{ M_1}(f)\le 1$ for any measurable
function $f$ with $\chi_{\iqw}\le f\le 1$; and similarly for $M_2$.

Conversely, assume that $M$ is such that for given
$\veps_1,\veps_2,\delta$, there exist $w,w'<\infty$ such that ($\alpha'$),
($\beta'$) hold. We show that ($\alpha$), ($\beta$) hold. It suffices to consider the
case of ($\alpha'$) implying ($\alpha$).

For each $q\in\R$, the  functions $f\in C_{uc}(\R)$ with
$\chi_{\iqw}\le f\le 1$ form a decreasingly directed set which
converges to $\chi_{\iqw}$. In fact, one can easily construct a
decreasing sequence of uniformly continuous functions $f_n$ with
$\chi_{\iqw}\le f_n\le 1$ and support in $[q-\delta/2-1/n,q+\delta+
1/n]$ that converges to $\chi_{\iqw}$. It follows that for every
$\rho$, the sequence of numbers $\rho^{ M_1}(f_n)\to \rho^{
M_1}(\iqw)$ as $n\to\infty$. (See \cite[Theorem 11.(iii)]{NOST}.)
Since for all $\rho$ with $\rho^\sfq(\iqd)=1$ we have $\rho^{
M_1}(f_n)\ge 1-\veps_1$, then also $\tr{\rho M_1(\iqw)}\ge 1-\veps_1$
for such $\rho$. 
\end{proof}

\begin{lemma}\label{lem:mav-approx}
Let $M$ be an $(\veps_1,\veps_2)$-approximate joint observable for
$\sfq,\sfp$. Then the covariant 
 linear map
 $\mav$ 
obtained from $M$
satisfies the conditions described in the preceding Lemma for the
given $\veps_1,\veps_2$.
\end{lemma}

\begin{proof}
It suffices to consider the statement for $\mav_1$,
that is: we show that for any $\veps_1\in(0,1)$, $\delta>0$, there
is a positive $w<\infty$ such that ($\alpha'$) holds for $\mav_1$.

Thus, given $\veps_1\in(0,1)$, $\delta>0$, there is $w<\infty$ such
that ($\alpha'$) holds for $M_1$. Now note that for $f\in C_{uc}(\R)$ with
$\chi_{\iqw}\le f\le 1$ the function $F$ on $\R^2$, defined
by $F(q,p)=f(q)$, is also uniformly continuous and satisfies
$\chi_{\iqw\times I}\le F\le 1$ and $M(F)=M_1(f)$.
Then the property ($\alpha'$) can be expressed equivalently as follows: for
all $q\in\R$, all $F\in C_{uc}(\R^2)$ with $\chi_{\iqw\times
\R}\le F\le 1$ and all $\rho$ with $\rho^\sfq(\iqd)=1$, we have
$\rho^{ M}(F)\ge 1-\veps_1$.

Consider the terms
\begin{equation*}\begin{split}
\tr{\rho M^{(q',p')}(F)}&=\tr{\rho W(q',p')^* M(\tau_{(q'p')}F)W(q',p')}\\
&=
\tr{W(q',p')\rho W(q',p')^*\,M(\tau_{(q'p')}F)}
\end{split}
\end{equation*}
for any state $\rho$, any $(q',p')\in\R^2$, and any $F\in
C_{uc}(\R^2)$. If $F$ runs through all such functions
satisfying $\chi_{\iqw\times \R}\le F\le 1$, and $\rho$ is any
state with $\rho^\sfq(\iqd)=1$, then $\tau_{(q',p')}F$ runs
through all uniformly continuous functions with the property
$\chi_{\tau_{(q',p')}\iqw\times \R}\le \tau_{(q',p')}F\le 1$,
and $W(q',p')\rho W(q',p')^*$ runs through all states localized in
$J_{q+q';\delta}$.

We can thus conclude that the functions $u(\rho,F)$ used in
(\ref{inv-mean}) to define $\mav$ satisfy $ u(\rho,F)(q',p')\ge
1-\veps_1$,
 and therefore ${\rm tr}\big[\rho \mav(F)\big]\ge 1-\veps_1$
for all uniformly  continuous $F$ with
$\chi_{\iqw\times \R}\le F\le 1$ and all $\rho$ localized in $\iqd$.   
\end{proof}

We will show that under the assumptions of Theorem 1 for $M$, which
are now seen to apply to $\mav$ in the form described in Lemma 2,
the functional $\mav$ extends to a normalized POM which is thus a
covariant phase space observable, and which inherits the property of
being an approximate joint measurement. According to \cite[Lemma
3]{Werner04b}, these results will follow if $\mav$ can be shown to have
zero weight at infinity.

The set of operators 
\[
\{\mav(f)\,:\,f\in C_{uc}(\R^2),\ f\ {\rm has\
compact\ support},\ 0\le f\le 1\}
\] 
forms an increasingly directed
net with upper bound $\mav(1)$, so that there is a supremum which we
denote $I-\mav(\infty)$. We have to show that $\mav(\infty)=O$, that
is, the supremum of the above set is the unit operator $I=\mav(1)$.
(This is the statement that the functional $\mav$ has zero weight at
infinity.) According to part 2 of Lemma 2 in \cite{Werner04b}, this
follows if one can show that $M_1(\infty)=M_2(\infty)=O$ (where
these operators are similarly defined).

\begin{lemma}
Let $M$ be an approximate joint observable for $\sfq,\sfp$, with
associated covariant $\mav$. Then the associated linear maps
 $\mav_1,\mav_2$ have zero weight at infinity, in the
following sense: for all $\rho\in S$,
\begin{equation}\label{sup1}
\sup\{\tr{\rho \mav_i(f)}\,:\,f\in C(\R),\ f\ {\rm has\ compact\ support},\ 0\le f\le 1\}=1.
\end{equation}
Thus $\mav_1(\infty)=\mav_2(\infty)=O$ and therefore $\mav(\infty)=O$.
\end{lemma}

\begin{proof}
It is sufficient to carry out the proof for $\mav_1$, using the fact
that $\mav$ is also an approximate joint observable.
Let $\rho$ be any state. Let $\veps_1\in(0,1)$ be given. We have to
show that there is a nonnegative function $f\in C_{uc}(\R)$, $0\le
f\le 1$, with compact support such that $\tr{\rho \mav_1(f)}\ge
1-\veps_1$.

We show this first for $\rho$ with $\rho^\sfq(\iqd)=1$ for some
$q,\delta$. In that case, given $\veps_1\in(0,1)$, there is a
positive finite $w$ and a function $f\in C_{uc}(\R)$ having compact
support with $\chi_{\iqw}\le f\le 1$ such that $\tr{\rho
\mav_1(f)}\ge 1-\veps_1$. Thus Eq.~(\ref{sup1}) holds.

Now consider any state $\rho$. Let $J_N=[-N,N]$, put
$\sfq_N=\sfq(J_N)$. Then, since $\sfq_N$ converges to $I$ ultraweakly,
we have eventually $\tr{\rho\sfq_N}\ne 0$, and we can define $\rho_N=\sfq_N\rho\sfq_N/\tr{\rho
\sfq_N}$. Then $\rho-\rho_N\to O$ in trace norm. (Write
$\sfq_N'=I-\sfq_N$ and
$\rho=\sfq_N\rho\sfq_N+\sfq_N'\rho\sfq_N'+\sfq_N\rho\sfq_N'
+\sfq_N'\rho\sfq_N$. For any effect $F$, we can estimate:
\[\begin{split}
|\tr{(\rho-\rho_N)F}| & \le \left|\frac 1{\tr{\rho \sfq_N}}-1\right|\tr{\sfq_N\rho\sfq_NF}+
|\tr{\sfq_N'\rho\sfq_N'F}|\\
&\qquad\qquad\qquad\qquad+|\tr{\sfq_N\rho\sfq_N'F}|+|\tr{\sfq_N'\rho\sfq_NF}| \\
& \le \left|\frac 1{\tr{\rho \sfq_N}}-1\right|\tr{\sfq_N\rho\sfq_N}+
|\tr{\sfq_N'\rho\sfq_N'}|\\
&\qquad\qquad\qquad\qquad+
2\left(\tr{F^2\sfq_N\rho\sfq_N}\right)^{1/2}\,\left(\tr{\rho\sfq_N'}\right)^{1/2}\\
& \le  2\tr{\rho\sfq_N'}+2\left(\tr{\rho\sfq_N'}\right)^{1/2}.
\end{split}
\]
In the second line we have used the Cauchy-Schwarz inequality for
Hilbert-Schmidt operators and $O\le F\le I$, and in the last line we
used $O\le F^2\le I$. All terms in the last line tend to 0 as
$N\to\infty$ (since $\tr{\rho \sfq_N'}\to 0$), and their sum is an
upper bound for the l.h.s. for all effects $F$. Since $\rho-\rho_N$
has zero trace, the trace norm is given by $\|\rho-\rho_N\|_{\rm
tr}=2\sup_{O\le F\le I}|\tr{(\rho-\rho_N)F}|$, and this tends to
zero as $N\to\infty$.)

Given $\veps_1\in(0,1)$ and $\rho\in S$, choose $N$ such that
$\|\rho-\rho_N\|_{\rm tr}\le \veps_1/2$. We know that for $\rho_N$
there is a uniformly continuous $f_N$ with $0\le f_N\le 1$ such that
$\tr{\rho_N\mav_1(f_N)}\ge 1-\veps_1/2)$. Then $\tr{\rho
\mav_1(f_N)}\ge\tr{\rho_N\mav_1(f_N)}-\veps_1/2\ge 1-\veps_1$.  
\end{proof}

We summarize the above considerations:

\begin{lemma}\label{lem:4}
Let $M$ be an approximate joint observable for $\sfq,\sfp$. The associated $\mav$ extends
to a covariant phase space observable of the form (\ref{gen-cov-obs}), denoted again $\mav$, and this
is in turn an approximate joint observable for $\sfq,\sfp$ with
\begin{equation}\label{mav-M-resol}\begin{split}
\delq{M_1}&\ge\delq{\mav_1}\ge\deq{\mav_1},\\
\delp{M_2}&\ge\delp{\mav_2}\ge\dep{\mav_2}.
\end{split}
\end{equation}
\end{lemma}

\begin{proof}
It remains to verify the inequalities, and here it suffices to show
that\break $\delq{M_1}\ge\delq{\mav_1}$.

Let $\veps_1\in(0,1),\delta>0$ be given. Let $w$ be a positive finite number such that for any
$q\in\R$ and all $\rho$ with $\rho^\sfq(\iqd)=1$, we have $\rho^{ M_1}(\iqw)\ge 1-\veps_1$. We conclude that for any $F\in C_{uc}(\R^2)$ with $\chi_{\iqw\times\R}\le F\le 1$, we obtain
$\rho^{ M}(F)\ge 1-\veps_1$,
and therefore, following the reasoning of the proof of Lemma \ref{lem:mav-approx},  also
$\rho^{ \mav}(F)\ge 1-\veps_1$. Since these functions $F$ form a decreasingly directed set converging
to $\chi_{\iqw\times\R}$, it follows also that $\rho^{\mav_1}(\iqw)\ge 1-\veps_1$.

So we have shown that $w\ge \delq{M_1}$ implies $w\ge\delq{\mav_1}$.
 \end{proof}

Since $\mav$ is a covariant phase space observable, Proposition \ref{cov-obs-ur}
applies and we have the measurement uncertainty relation (\ref{cov-error-bar-ur})
for $\mav$. The inequalities (\ref{mav-M-resol}) finally yield the general uncertainty relation
for error bars (\ref{error-bar-ur}).

The inequality (\ref{res-ur}) for resolution widths follows similarly as a consequence of the
inequalities
\begin{equation}
\gamma_{\veps_1}(M_1)\ge \gamma_{\veps_1}(\mav_1),\quad
\gamma_{\veps_2}(M_2)\ge \gamma_{\veps_2}(\mav_2),
\end{equation}
the proof of which is analogous to the argument in the proof of
Lemma \ref{lem:4} and will thus be omitted. Theorem 1 is thus
proven.

An investigation of the scope and applications of this result will
be given elsewhere \cite{BuPe07}. Here we conclude with a comparison
of the present approach with that of R.~Werner \cite{Werner04b} from
which we have adopted the proof strategy for our Theorem 1. Werner
defines a distance  $d(E_1,E_2)$ on the set of observables on $\R$
as follows.

First recall that for any bounded measurable
function $h:\R\to\R$, the integral $\int_\R h\,dE$ defines (in the weak
sense) a bounded selfadjoint operator, which we denote by $E[h]$.
Thus, for any vector state $\fii$ the number $\ip{\fii}{E[h])\fii}=\int_\R h\,d\prob^E_\fii$ is well-defined.

Denoting by $\Lambda$ the set of bounded measurable functions $h:\R\to\R$ for which $|h(x)-h(y)|\leq |x-y|$, the distance between the observables $E_1$ and $E_2$ is  defined as
\begin{equation}\label{Werner-distance}
d(E_1,E_2) := \sup_{\rho\in S}\ \sup_{h\in\Lambda}\
\left|\tr{\rho  E_1[h]}-\tr{\rho E_2[h] }  \right|.
\end{equation}

Werner proved the following joint-measurement uncertainty relation, valid for any
observable $M$ on phase space with marginals $M_1,M_2$:
\begin{equation}\label{W-ur}
d(M_1,\sfq)\cdot d(M_2,\sfp)\ge C\hbar.
\end{equation}
The tightest lower bound for the product of distances can be
determined  within the class of covariant  phase space observables
and has a value of approximately 0.3047.

We show that the condition of finite distance is stricter than that
of finite error bar width.

\begin{proposition}\label{finite-dist-approx}
Any observable $E_1$ on $\R$  that satisfies the condition
$d(E_1,E)<\infty$ for a sharp observable $E$ on $\R$ is an
approximation to $E$ in the sense of finite error bars. In that case
the following inequality holds:
\begin{equation}\label{we-leq-d}
\dee{E_1}\le\frac 2\veps\,d(E_1,E).
\end{equation}
\end{proposition}

\begin{proof}
 We are given that
\[
\big|\tr{\rho E_1(h)}-\tr{\rho E(h)}\big|\le d(E_1,E)=:c\quad{\rm for\ all\ }\rho\in S,\ h\in\Lambda.
\qquad(+)
\]
Let $\veps\in(0,1)$ and $\delta>0$ be given. Put
$w=\delta+2n$, with $n\in\N,\ n\ge c/\veps$. 
Consider an interval $\iqd$ and a state $\rho$ with $\rho^E(\iqd)=1$. Define
the functions $h_n$ via
\[
h_n(x):=\left\{\begin{array}{ll} n&{\rm if}\ \ |x-q|\le\delta/2;\\
n+\delta/2-|x-q|&{\rm if}\ \ \delta/2<|x-q|\le \delta/2+n;\\
0&{\rm if}\ \ \delta/2+n<|x-q|.\\
\end{array}
\right.
\]
Note that $h_n\in\Lambda$. Condition $(+)$ for $h_n$ entails for
$g_n=h_n/n$ that $\big|\rho^{E_1}(g_n)-\rho^E(g_n)\big|\le c/n$. We
then have 
$\chi_{\iqd}\le g_n\le \chi_{\iqw}$.

Now $\rho^E(\iqd)=1$ implies $\tr{\rho E(g_n)}=1$, and so, using the assumption
$n\ge c/\veps$,  we obtain
\[
\tr{\rho E_1(\iqw)}\ge\tr{\rho E_1(g_n)}\ge\tr{\rho E(g_n)}-c/n\ge 1-\veps.
\]
To prove the inequality (\ref{we-leq-d}), we note that on putting
$w=\delta+2c/\veps$, one still obtains $\tr{\rho E_1(\iqw)}\ge
1-\veps$. This yields $\dele{E_1}\le\delta+2d(E_1,E)/\veps$, and on
letting $\delta$ approach 0, then (\ref{we-leq-d}) follows.
\end{proof}

An immediate consequence of Eqs. (\ref{we-leq-d}) and
(\ref{qp-err-res}) for an approximate position observable $\sfq_\mu$
is the following:
\begin{equation}
W_{\veps_1}(\mu)=\gamma_{\veps_1}(\sfq_\mu)\le\deq{\sfq_\mu}\le
\frac 2{\veps_1}\, d(\sfq_\mu,\sfq).
\end{equation}
This gives a bound on the resolution width of $\sfq_\mu$ and on the
overall width of the unsharpness measure $\mu$, showing the
behaviour of these quantities as $\veps_1\to 0$.

There are instances of joint measurements for which Werner's distances are
infinite while the error bar widths are finite. This can be seen in the case of covariant
phase space observables where the relevant distance between (say) the marginal
$\sfq_{\mu_\bom}$ and $\sfq$ is $d(\sfq_{\mu_\bom},\sfq)=\int |q|\mu_\bom(dq)$ (see
\cite{Werner04b}).

Finally we note that there exist non-covariant observables on phase
space which are approximate joint observables for $\sfq$ and $\sfp$.
An example is $M:=G^{\bom}\circ \gamma^{-1}$, where
$\gamma=(\gamma_1,\gamma_2)$ is a bijective measurable map of $\R^2$
onto itself; $M$ is an approximate joint observable if
$\gamma_1(q)-q$ and $\gamma_2(p)-p$ are bounded functions, and $M$
is non-covariant if $\gamma_1$ or $\gamma_2$  is not an affine map
(see \cite{BuPe07} for details).

\section{Conclusion}

We have introduced an operationally significant and experimentally relevant criterion,
based on the new concept of error bar width,
of what constitutes an approximate joint observable of position and momentum.
The associated error bar widths obey a Heisenberg uncertainty relation. This shows that
the approximations of position and momentum in terms of marginals of an observable
on phase space cannot both be arbitrarily good.

We also considered the resolution width as an indicator of the degree of intrinsic unsharpness.
It was found that the resolution widths of the marginals of any approximate joint observable for position and momentum cannot both be arbitrarily small but must obey a Heisenberg uncertainty relation.

\vspace{12pt} 

\noindent {\bf Acknowledgement.}
The authors would like to thank Pekka Lahti 
and Werner Stulpe
for valuable comments on an earlier manuscript version of this paper.



\end{document}